\documentstyle[12pt]{article}
\begin{document}

\begin{titlepage}
\nopagebreak
\begin{flushright}

hepth@xxx/9207027
\\
LPTENS--92/20
\\
ENSLAPP-L-393/92
\\
July  1992
\end{flushright}

\vglue 2.5  true cm
\begin{center}
{\large\bf
SUPERSYMMETRIC BLACK HOLES \\ FROM \\TODA THEORIES\footnote{
\LaTeX \ file  available from
hepth@xxx.lanl.gov (\# 9207027)}}

\vglue 1  true cm
{\bf Fran\c cois DELDUC}\\
{\footnotesize Laboratoire de Physique Th\'eorique de
l'ENS Lyon, \\ 46 All\'ee  d'Italie, 69364 Lyon C\'EDEX 07, France.}\\
{\bf Jean-Loup~GERVAIS}\\
{\footnotesize Laboratoire de Physique Th\'eorique de
l'\'Ecole Normale Sup\'erieure\footnote{Unit\'e Propre du
Centre National de la Recherche Scientifique,
associ\'ee \`a l'\'Ecole Normale Sup\'erieure et \`a l'Universit\'e
de Paris-Sud.},\\
24 rue Lhomond, 75231 Paris C\'EDEX 05, ~France.
\\and\\
}
{\bf Mikhail V.  SAVELIEV}\footnote{ On leave of absence from
the Institute for High Energy Physics,
142284, Protvino, Moscow region, Russia.}\\
{\footnotesize Laboratoire de Physique Th\'eorique de
l'ENS Lyon, \\ 46 All\'ee  d'Italie, 69364 Lyon C\'EDEX 07, France.}\\
\medskip
\end{center}

\vfill
\begin{abstract}
\baselineskip .4 true cm
\noindent
By the example of nonabelian Toda type theory associated with the
Lie superalgebra $osp(2|4)$ we show that this integrable
dynamical system is relevant to a black hole background
metric in the corresponding target space.
In the even sector the model under consideration reduces to the exactly
solvable conformal theory (nonabelian $B_2$ Toda system)
in the presence of a black hole recently proposed in \cite{1}.

\end{abstract}
\vfill
\end{titlepage}
\baselineskip .5 true cm

{}1. It was shown in paper \cite{1} that  exactly solvable
nonabelian Toda systems \cite{2} are relevant
to two dimensional conformal field theories
in the presence of a black hole. Here the tachyon fields
arising as the potential terms in the
corresponding Lagrangian, play the role of
the 2d cosmological terms. Moreover, the corresponding
$B_n$-WZNW model is gauged by an appropriate nilpotent
group, in distinction with those of Witten's scheme,
see \cite{3} and references therein.
The simplest nontrivial example containing
all main features characteristic of  the black
holes in the framework of this approach, is based
on the Lie algebra $B_2$ supplied with a noncanonical gradation.

In the present note we discuss a supersymmetric
black hole by the example of
$N=1$ supersymmetric extension of nonabelian
Toda system associated with the
classical finite dimensional Lie superalgebra
$osp(2|4)$. Here we use such a choice of its simple roots that
 in the corresponding Dynkin diagram the first simple
 root is odd (it corresponds to $sl(1|1)$ superalgebra)
 and other two simple roots are even. With an appropriate
 gradation, this superalgebra provides,   in the even sector,
 the reduction of the system under consideration to the case
 of the Lie algebra $B_2$.

{}2. Consider the classical Lie superalgebra ${\cal G}$ with the
defining relations,\\ see e.g., \cite{4},
\begin{equation}
[h_i,h_j]=0,\quad[h_i,X_{\pm j}]=\,\pm k_{ji}X_{\pm j},
\quad [X_{+i},X_{-j}]=\delta_{ij}h_i,
\end{equation}
where $k$ is the Cartan matrix of ${\cal G}$; $h_i$ and $X_{\pm i},\,1
\leq i\leq $ rank ${\cal G}$, are its Cartan and Chevalley generators.
Here
the numerical indices of the elements of the superalgebra ${\cal G}$
 correspond to its  roots, in particular,
subscripts of the generators denote the simple roots. For the case
under consideration $k$ is defined in accordance  with
the above mentioned Dynkin scheme. Realize the gradation of
the $osp(2|4)$ by the Cartan element $H=h_2+2h_3$ of its $osp(1|2)$
subsuperalgebra with the odd elements $Y_{\pm}=X_{\mp 1}
 \mp X_{\pm 123}$, and even elements (generators of the $sl(2,C)$
 subalgebra of $osp(1|2)$): $H=[Y_+, Y_-]_+, X_{\pm}=Y_{\pm}^2$.
Then the subspaces ${\cal G}_m, m=0, \pm 1, \pm 2,$ of $osp(2|4)=
\oplus _m {\cal G}_m$, are the following:
\begin{eqnarray}
{\cal G}_0 & = & \{X_{\pm 2}, h_i, ,1 \leq i\leq 3 \};\\
{\cal G}_{\pm 1} & = & \{X_{\mp 1}, X_{\mp 12}, X_{\pm 123},
X_{\pm 1223}\};\\
{\cal G}_{\pm 2} & = & \{X_{\pm 3}, X_{\pm 23}, X_{\pm 223}\}.
\end{eqnarray}

Start with the $N=1$ superextension \cite{5} of the Toda systems
(Abelian as well as their nonabelian versions) associated with finite
dimensional classical Lie superalgebras  ${\cal G}$,
\begin{equation}
D_+(g^{-1}D_-g)=[Y_-, g^{-1}Y_+g]_+,
\end{equation}
where $g$ is a regular element of the Grassmann span of the
Lie group $G_0$
with Lie algebra ${\cal G}_0$, and depends on the coordinates of the $2|2$
superspace ${\cal C}_{2|2}$; $D_{\pm}$ are the supersymmetric
covariant derivatives in this superspace. Then, in the case under
 consideration,
parametrize
the function $g$ as
\begin{equation}
g=\exp \{a^+ X_{+2} \} \;\exp \{a^- X_{-2} \}\;\exp
\{\,\sum_{i=1}^{3}a_i h_i\},
\end{equation}
with the unknown functions (superfields) depending on the points in
${\cal C}_{2|2}$. We come to the following system of equations:
\begin{eqnarray}
& & D_+D_-a_1=e^{-a_2}-e^{-a_1+a_2-a_3}(1+a^+a^-),\nonumber\\
& & D_+[D_-a_2+a^-D_-a^+]=e^{-a_1+a_2-a_3}(1+a^+a^-),\nonumber\\
& & D_+D_-a_3=2e^{-a_1+a_2-a_3}(1+a^+a^-),\nonumber\\
& & D_+[e^{-a_1-2a_2+a_3}D_-a^+]=-e^{-2a_1-a_2}a^+,\nonumber\\
& & D_+[e^{a_1+2a_2-a_3}(D_-a^--(a^-)^2D_-a^+)]=-e^{a_1+a_2-a_3}a^-.
\end{eqnarray}

Now, it is the very time to note that there is such an element
$H_0\equiv h_1 - h_3$ which commutes with $Y_{\pm}$, and,
consequently, with $X_{\pm}$, while $h_2$ commutes only
with $X_{\pm}$, and just due to this reason the r.h.s. of (5) in
our case contains only $h_1$ and
$H$ but not $h_2$ (as it should be in the even case). Then, making use the
gauge
properties of our system, equations (7) can be transformed, in
 terms of the functions
\[
\rho \equiv -\mbox{Ar}\mbox{sh}(a^+a^-)^{1/2}-i\frac{\pi}{2}; \quad
\psi\equiv -\frac{a_1+a_3}{2}+i\frac{\pi}{2};
\]
and
\[
D_+\omega =-\frac{1}{2(1+a^+a^-)}[a^-D_+a^+-(1+2a^+a^-)a^+D_+a^-
\]
\[
-2a^+a^-(1+a^+a^-)D_+(a_1+2a_2-a_3)],
\]
\[
D_-\omega =\frac{1}{2(1+a^+a^-)}[-a^+D_-a^-+(1+2a^+a^-)a^-D_-a^+];
\]
to the form
\begin{eqnarray}
& & D_+D_-\rho  =  \mbox{ch}\rho\,e^{\psi}-
\frac{\mbox{sh}\rho}{\mbox{ch}^3\rho}
\,D_+\omega \, D_-\omega,\nonumber\\
& & D_+(\mbox{th}^2\rho  D_-\omega)-D_-(\mbox{th}^2\rho \,D_+\omega)
=0,\nonumber\\
& &  D_+D_-\psi  =   \mbox{sh}\rho\,e^{\psi}.
\end{eqnarray}

The Lagrangian density ${\cal L}$ corresponding to the system (8)
is the following:
\begin{equation}
{\cal L} = \frac{1}{2}D_+\rho\,D_-\rho+\frac{1}{2}D_+
\psi\,D_-\psi - \frac{1}{2}\mbox{th}^2 \rho D_+\omega\,D_-\omega +
\mbox{sh}\rho\,e^{\psi}.
\end{equation}
Therefore, the target-space metric $G$,
\begin{equation}
G_{ij}=\mbox{ diag } (1, 1, -\mbox{ th }^2\rho),
\end{equation}
contains the hyperbolic-tangent-square function (superfield)
which, in even case, is characteristic for the black hole
 in the spirit of Witten. The potential term in (9) plays
 the role of the cosmological term in
${\cal C}_{2|2}$.

We are not going to discuss here more details concerning the
properties of supersymmetric black holes in the framework
of the given approach; only note
once more that they are described by exactly solvable system of equations.
A relation with the corresponding gauged supersymmetric WZNW
model is quite evident.

{\bf Acknowledgements.} One of the authors (M.S.) would like to
thank A. Izergin, A. Sevrin, and P. Sorba for useful
discussions; and ENSLAPP in Lyon for kind hospitality.

\end{document}